\documentclass[superscriptaddress,floatfix,twocolumn,amsmath,amssymb]{revtex4-1}

\usepackage{graphicx}
\usepackage{dcolumn}
\usepackage{bm}
\usepackage{amssymb}
\usepackage{natbib}
\usepackage{mathtools}
\usepackage{color}
\usepackage{datetime}
\usepackage{footnote}
\usepackage{bbold}
\usepackage[normalem]{ulem}
\usepackage{dcolumn}
 \usepackage{ragged2e}
 \usepackage{xfrac}
 \usepackage{sidecap}
 \usepackage{setspace}
 \usepackage{multirow}
\usepackage{xcolor}
\usepackage{hyperref}
\usepackage{cleveref}
\usepackage{microtype}
\usepackage{siunitx}
\usepackage[utf8]{inputenc}
\newcommand{\ngr}{n_\mathit{g}}
\newcommand{\ngrte}{n_\mathit{g}^\mathrm{TE}}
\newcommand{\ngrtm}{n_\mathit{g}^\mathrm{TM}}
\newcommand{\um}{$\mu$m }

\begin{document}

\title{Tunable large free spectral range microring resonators in lithium niobate on insulator}

\author{Inna Krasnokutska}
\thanks{These authors contributed equally to this work}
\affiliation{Quantum Photonics Laboratory and Centre for Quantum Computation and Communication Technology, School of Engineering, RMIT University, Melbourne, Victoria 3000, Australia}

\author{Jean-Luc J. Tambasco}
\thanks{These authors contributed equally to this work}
\affiliation{Quantum Photonics Laboratory and Centre for Quantum Computation and Communication Technology, School of Engineering, RMIT University, Melbourne, Victoria 3000, Australia}

\author{Alberto Peruzzo}
\thanks{alberto.peruzzo@rmit.edu.au}
\affiliation{Quantum Photonics Laboratory and Centre for Quantum Computation and Communication Technology, School of Engineering, RMIT University, Melbourne, Victoria 3000, Australia}

\begin{abstract}
Microring resonators are critical photonic components used in filtering, sensing and nonlinear applications.
To date, the development of high performance microring resonators in LNOI has been limited by the sidewall angle, roughness and etch depth of fabricated rib waveguides.
We present large free spectral range microring resonators patterned via electron beam lithography in high-index contrast $Z$-cut LNOI.
Our microring resonators achieve an FSR greater than 5~nm for ring radius of 30~$\mu$m and a large 3 dB resonance bandwidth.
We demonstrate 3~pm/V electro-optic tuning of a 70~$\mu$m-radius ring.
This work will enable efficient on-chip filtering in LNOI and precede future, more complex,
microring resonator networks and nonlinear field enhancement applications.
\end{abstract}

\maketitle

\section{Introduction}
Microring resonators are fundamental components in any high-index contrast photonic platform \cite{Bogaerts:2012,Vahala:03}.
They are a highly sought after cavity component, as they enable on-chip field enhancement as well as
spectral filtering and fast modulation of optical signals \cite{Bogaerts:2012, Xia:07, Baba:13, Levy:11, Hu:12}.
In the past decade, microring resonators have been demonstrated in a multitude of platforms including silicon (Si) \cite{Xu:08, Shi:12}, silicon nitride (SiN) \cite{Popovic:06}, aluminium nitride (AlN) \cite{Pernice:12, Pernice2:12}, galium arsenide (GaAs) \cite{Absil:00, Ibrahim:02} and indium phospide (InP) \cite{Ciminelli:13}. The applications of microring resonators are vast, ranging from sensing biological samples \cite{Guo:06}, to filtering and demultiplexing telecommunication lines \cite{Xia:07, Ibrahim:02}, and generating frequency combs for spectroscopy \cite{Jung:13}.

Microring resonators are challenging photonic components to fabricate, as losses incurred in the cavity are greatly amplified. To achieve a large free spectral range (FSR) for telecommunication applications and sensing, small, single-mode high-index contrast waveguides are required. Microring resonators can also be cascaded to increase the spectral enhancement, or create various types of filters \cite{Kim:16} and this requires very precise and careful control of the 3 dB resonance bandwidth and FSR \cite{Popovic:06}. In general, a ring performance is limited either by the material properties, such as 2-photon absorption in the C-band of Si, or the ability to nanostructure the material to produce small waveguides with smooth sidewalls.
Lithium Niobate (LN) could greatly benefit from microring resonators to enhance its nonlinear and electro-optic (EO) properties, as well as prepare it for telecommunication use. Traditionally, LN has only supported prohibitive low-index contrast waveguides made from Ti:LN \cite{Janner:2009} and PE:LN \cite{Jackel:82, Stepanenko:16}. 
With the commercialization of Lithium Niobate On Insulator wafers (LNOI) \cite{Poberaj:2012kf}, high-index contrast waveguides in LN can now be achieved.
Small radius, multimode waveguide rings have been reported in LNOI, but suffered from high propagation loss due to fabrication imperfection \cite{Poberaj:2012kf, 7353164}. Recent improvement in the fabrication process has led to reduced sidewall roughness, enabling the fabrication of ultra-low loss waveguides \cite{Zhang:17,Krasnokutska:18} and microring resonators with extremely high $Q$-factors \cite{Zhang:17}. Due to the complex nature of processing LNOI, low-loss, single-mode, compact rings (down to 30\,\um of radius) in LNOI with a high FSR are yet to be reported. Due to the complex chemistry of etching LNOI, most etching processes result in $\sim50^\circ$ sidewall angle waveguide \cite{Wang:2017ul, Liang:17}, hampering the ability to produce small gaps, which are critical in the fabrication of grating couplers, compact directional couplers and multistage microring resonator filters; however, this problem was recently solved where $\sim75^\circ$ sidewall angle waveguides were reported \cite{Krasnokutska:18}.

The tuning and reconfigurability of photonic components is a necessity for many practical applications.
Tunable rings resonators have been reported in several photonic platforms, including Si, which achieve a high-speed modulation via carrier-depletion and thermal wavelength tuning via resistive heaters \cite{Li:11}.  However, carrier depletion suffers from increased optical absorption and a limited response time, restricting the performance of high-speed switches.
Electro-optics offers a solution to these challenges, and has been demonstrated in Si on LN \cite{Chen:13, Chen:14}, AlN \cite{Jung:14} and LNOI \cite{Wang:18, Guarino:2007bv, Siew:cx}.
The devices reported to date in $Z$-cut LNOI have either required two-step laser lithography \cite{Guarino:2007bv} or had performance challenges due to the waveguides being multimode and having limiting propagation losses \cite{Siew:cx}.

In this work, we present a detailed study of all-pass microring resonators fabricated monolithically in $Z$-cut LNOI from small, low-loss, high-index contrast and single mode C-band waveguides. We analyze the performance of multiple rings with varying radii from 30 \um to 90 $\mu$m.  The demonstrated heavily overcoupled microring resonators have a maximum FSR of 5.7 nm and a large 3\,dB resonance bandwidth that both agree well with the design and simulation.  
In contrast to previous work who focused on reaching high Qs, this work aims at filtering applications where strong coupling between the ring and bus-waveguide is desired which results in a larger bandwidth of the cavity resonance and a reduced Q correspondingly.
Furthermore, we demonstrate the versatility of our fabrication process, etching down to 300\,nm trenches in LNOI, critical for advanced photonic components.  
We further report 3~pm/V electro-optic tuning of a 70~$\mu$m-radius microring resonator---to the authors' knowledge, this is the largest to date in $Z$-cut LNOI.
We expect the microring resonators in this work to pave the way towards on-chip filtering in LNOI with ring networks, as well as field enhancement applications such as switching and nonlinear photon generation.

\begin{figure*}[!ht]
\centering
\includegraphics[width=0.8\linewidth]{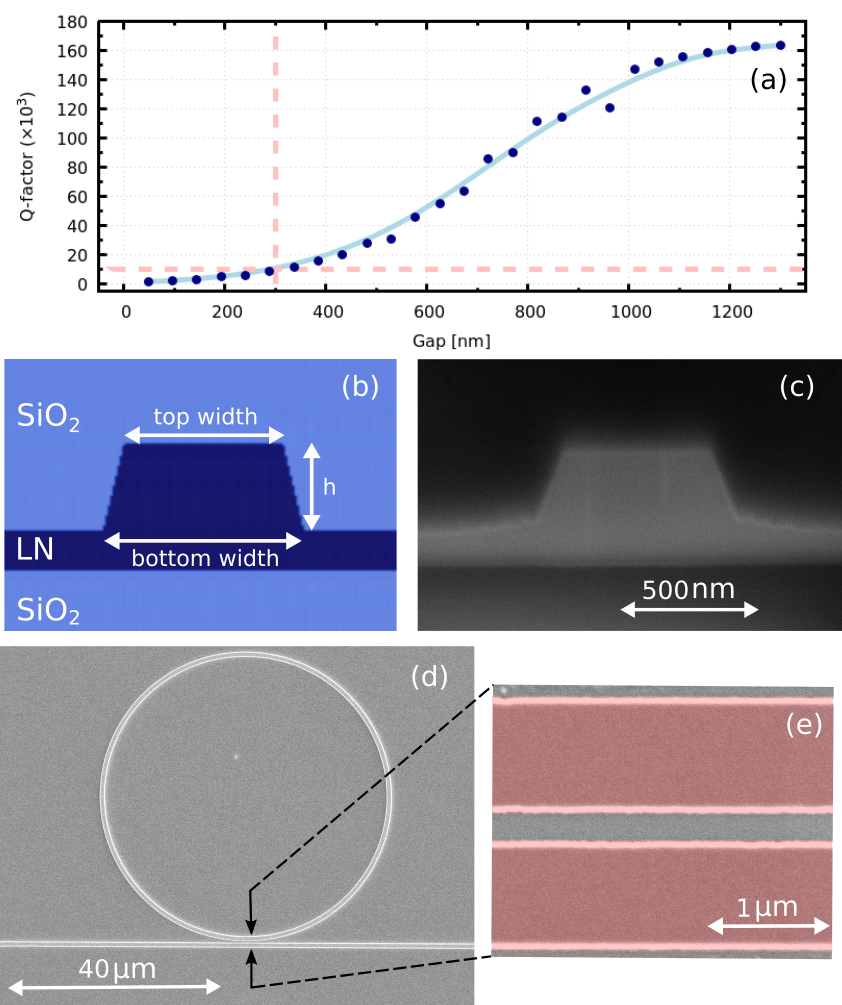}
\caption{(a) Simulation of $Q$-factor as a function of the coupling region gap for a 30 $\mu$m radius microring---the light red dashed lines demarcate the simulated $Q$-factor ($\sim$10000) for the microring shown in (d). (b) Design of a single mode LNOI rib waveguide at 1550 nm wavelength; the top width is 650 nm, the bottom width is 840 nm and the waveguide height is 350 nm. Scanning electron microscope pictures: (c) cross-section taken by FIB slicing and SEM imaging; (d) an etched 30 $\mu$m radius ring with a 300 nm gap between the bus waveguide and the ring prior to PECVD SiO$_{2}$ cladding; (e) false-color image of the coupling region after the lift-off process and prior to etching; the false-red highlights the metal etch mask.}
\label{fig1}
\end{figure*}

\section{Design and fabrication}
 Microring resonators with radii 30--90 \um were designed to obtain an FSR from 1.5 to 5.7 nm and were simulated using the commercially available software, Lumerical.  Rings of varying radii were fabricated to analyze the FSR and performance for the TE and TM modes. The small bending loss needed for good operation of a 30 \um microring resonator required high-index contrast single mode waveguides at 1550 nm. A mode solver was used to determine the dimensions required to ensure a sufficiently small TM polarization bend radius. The design of the waveguide includes the following parameters: rib height, top width, sidewall angle, refractive indices of the waveguide and claddings, and film thickness. The cross-section of a $Z$-cut rib waveguide cladded with SiO$_{2}$ is shown in Fig. \ref{fig1}(b).  The small gap of 300 nm was chosen to heavily overcouple the rings and obtain wide bandwidth resonances, rather than extremely narrow resonances that require very precise wavelength tuning to access. A simulation of the $Q$-factor as a function of coupling region gap for the TM mode at 1550 nm of a 30 \um microring resonator is shown in Fig. \ref{fig1}(a), and indicates that the 3\,dB bandwidth of the microring resonances (FWHM) is $\mathrm{FWHM}=\lambda_\mathrm{res}/Q=\sim$155\,pm, where $\lambda_\mathrm{res}$ is the wavelength of the resonance and $Q$ is the $Q$-factor.  The simulation was performed using Lumerical Mode; the measured losses, as per the Fabry-Perot measurements presented in Fig. \ref{fig2}(a), were taken into account in the simulation model.

The photonic components were fabricated by the process developed and described in our previous work \cite{Krasnokutska:18}. The process starts with 500 nm thick LN film, which is fabricated using the smart-cut technique on 2 \um of SiO$_2$ layer and supported by a 500 \um LN substrate. The next fabrication steps rely on electron beam lithography and lift-off of the e-beam evaporated metal layer to obtain a hard mask defining the photonic components. The scanning electron microscopy image (SEM) of a waveguide to a ring coupling region just after the metal lift-off process, is shown in Fig. \ref{fig1}(e). The components were then dry etched in a reactive ion etcher Fig. \ref{fig1}(d). Following etching, the waveguides were cladded with 3 \um thick plasma-enhanced chemical vapor deposition (PECVD) SiO$_{2}$. The presented structures were etched deeper than in our previous work to achieve the necessary index contrast, reducing the waveguide bending radius. The rib waveguide cross-section, obtained via focused ion beam (FIB) slicing and scanning electron microscopy (SEM), shows a sidewall angle of 75$^{\circ}$ and an etch depth of 350 nm Fig. \ref{fig1}(c). Finally, the waveguide facets were diced using optical grade dicing to facilitate butt-coupling. The length of the chip, after all processing steps were completed, is 6 mm.

\section{Experimental results}
\begin{figure*}[!ht]
\centering
\includegraphics[width=0.8\linewidth]{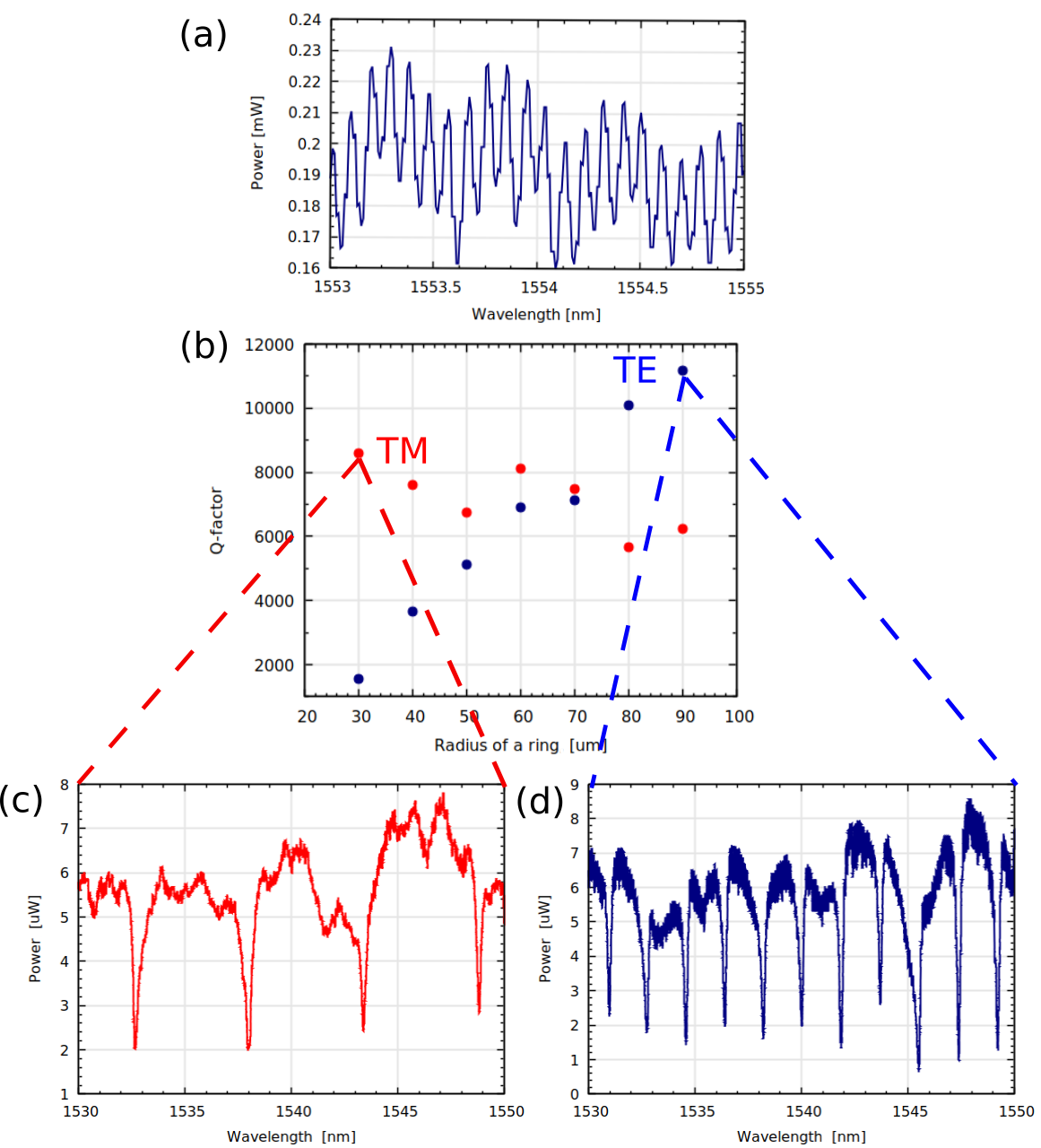}
\caption{(a) Transmission spectrum of the inverse taper with 200 nm width LNOI used for calculating the propagation loss of the TM mode.(b) Measured $Q$-factors of the ring resonators versus their radius for the TE and TM modes.  The blue curve corresponds to the TE mode and the red curve corresponds to the TM mode. (c) Spectral response of the ring with radius 30 \um for TM mode. (d) Spectral response of the ring with radius 90 \um for TE mode.}
\label{fig2}
\end{figure*}

In order to confirm that the photonic components are not limited by the propagation loss, loss measurements were performed prior to the characterization of the microrings, using the Fabry-Perot loss measurement technique \cite{Regener1985}. Laser light at 1550 nm wavelength is coupled into and out of the polished facets of the waveguide using polarization maintaining (PM) lensed fibers with a mode field diameter of 2 \um. A typical optical transmission spectrum for TM (the TE and TM modes have a similar response) is shown on the Fig. \ref{fig2} (a). Linear inverse tapers 200 \um long down to 200 nm width, at the waveguide ends, are used to improve the mode matching between the lensed fibre and the waveguide \cite{6895281}, and improve the signal to noise ratio of the Fabry-Perot measurements. As the waveguide narrows, the mode field diameter at the input and output of the waveguide significantly increases, allowing improved mode matching with the mode of the lensed fiber. The total input and output coupling and propagation loss is 8 dB for a 6 mm long chip, compared to the 15 dB loss achieved with the straight waveguide without tapering section. The estimated propagation loss is less than 0.5 dB/cm for both the TE and TM modes, which is in agreement with the results obtained in our previous work \cite{Krasnokutska:18}

The fabricated microring resonators were characterized by sweeping the wavelength of the laser between 1530 to 1610 nm and recording their spectral responses with a commercially available high-speed InGaAs photodiode. The laser light was injected into and out of a 6 mm bus-waveguide via PM lensed fibers. To decrease the chance of interference between multiple oscillations inside of the photonic component, the inverse tapering section was not implemented for the microrings---this led to a drop in the mode matching efficiency. We observe that both TE and TM (Fig. \ref{fig2}(b)) modes reaches the largest FSR for the ring with the smallest radius 30 \um; however, the TE and TM modes show different results in terms of the achievable $Q$-factor for this geometry. A $Q$-factor of $\sim9000$ was achieved for the TM mode whilst for the TE mode the $Q$-factor is significantly smaller $\sim1200$. As the radius of the ring increases, the $Q$-factor for TM mode remains almost unchanged (Fig. \ref{fig2}(b)); meanwhile, for the TE mode, it significantly increases (Fig. \ref{fig2}(b)) and the highest $Q$-factor has been achieved for the ring with radius of 90 \um Fig. \ref{fig2}(d). This dissimilarity can be attributed to the difference in the bending loss between both modes. It was deduced by using our numerical model (Fig. \ref{fig1}(a)) and the value of intrinsic quality factor that TM mode bending loss for microring resonator of 30 \um of radius is around 1.5 dB/cm, while for TE mode it estimated to be around 12 dB/cm for the ring with the same radius. By comparing theoretical and experimental results, the effective index for the TE mode is 1.85 and for the TM mode is 1.72 and the TM mode was confined to have an index contrast of $\sim0.272$, whilst the TE mode is lower $\sim0.247$.  As the TM mode has a higher index contrast, a smaller bend radius is achieved, enabling smaller microring resonators to be realized. The TE mode bending loss decreased with increasing microring resonator radius, leading to an improvement in the $Q$-factor.

\begin{figure*}[!ht]
\includegraphics[width=1\linewidth]{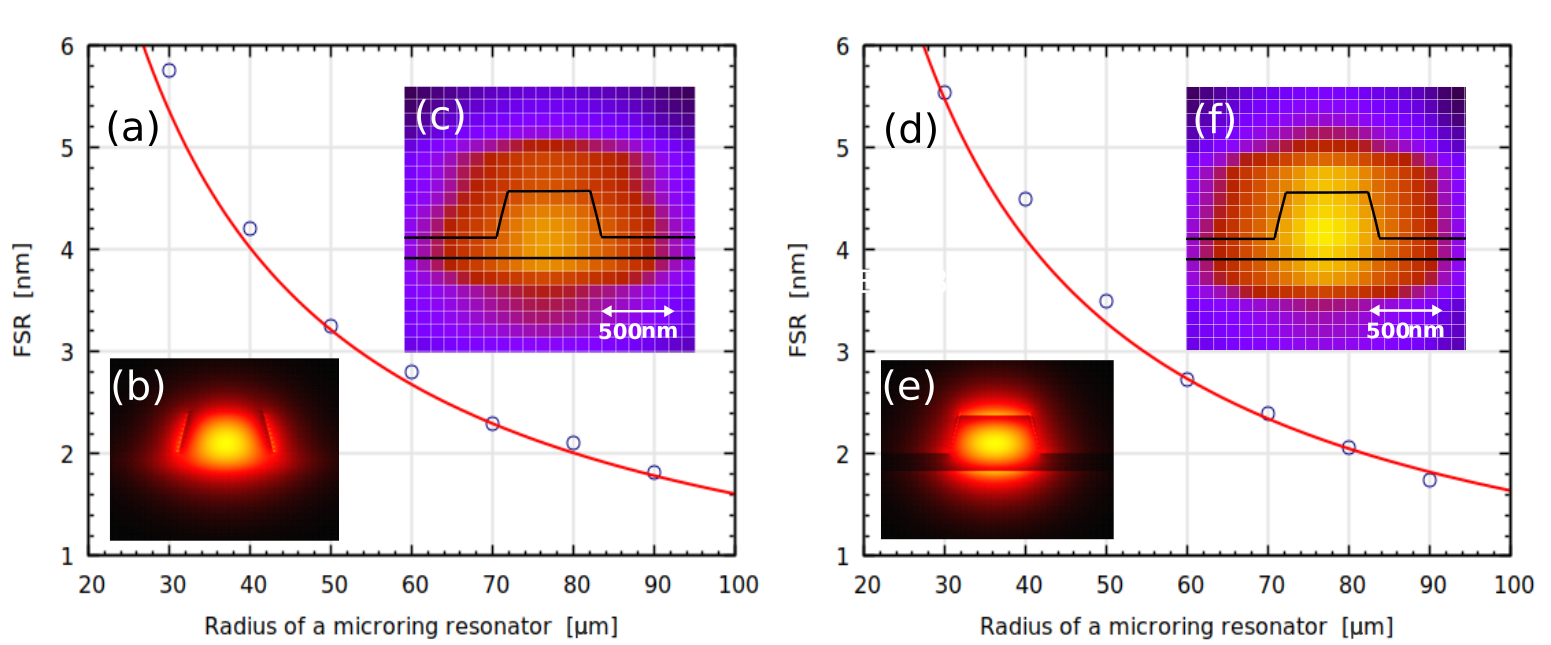}
\caption{(a) Measured FSR as the function of a microring resonator radius for the TE mode and (d) for TM mode; the blue circles are measured values, whilst red line is theoretically predicted dependence of FSR on microring resonator radius; (b) the simulated electrical field distribution for the TE mode and (e) for the TM mode; (c) measured optical power distribution at the output of the chip for the TE mode and (f) for the TM mode, where black lines schematically show the actual waveguide dimensions.}
\label{fig3}
\end{figure*}

The group indices for the TE and TM modes respectively, $\ngrte$ and $\ngrtm$, are deduced from the fully-vectorial mode solver using the Sellmeier equations for lithium niobate: $\ngrtm=2.33$ and $\ngrte=2.38$.  The FSR can be calculated using $\mathrm{FSR}=\lambda^2/(\ngr L)$, where $L$ is the circumference of the ring ($L=2\pi R$), $R$ is the radius of the ring. The simulation curve is plotted with the measured FSR for different microring resonator dimensions in Fig. \ref{fig3}(a) and Fig. \ref{fig3}(d).  The simulated $E$-field distributions of the fundamental waveguide modes at a wavelength of 1550 nm (found using an in-house mode solver) are included to the figures as insets: Fig. \ref{fig3}(b) for the TE mode, and Fig. \ref{fig3}(e) for the TM mode. Also included as insets, Fig. \ref{fig3}(c) and Fig. \ref{fig3}(f), show the measured power distribution at a wavelength of 1550 nm in a $3\times3$ \um window; each cell defined by the white grid lines represents a single pixel (a single power measurement).  The measured power distribution is performed by sweeping the fiber over the output facet of the waveguide, resulting in a convolution between the fiber mode and the waveguide mode, smearing and enlarging the appearance of the waveguide mode.

\section{Electro-optic resonant wavelength tuning}

\begin{figure*}[!ht]
\centering
\includegraphics[width=1\linewidth]{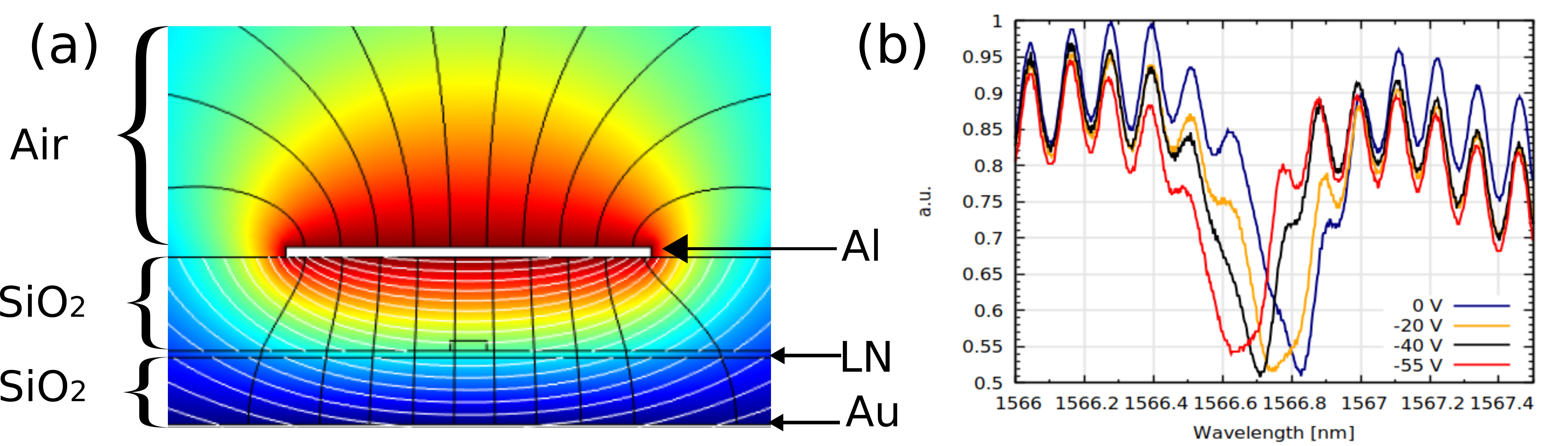}
\caption{(a) Simulation results of electrical field. b) Spectrum of the optical resonances when voltage from -55 to 0 V is applied.}
\label{fig4}
\end{figure*}

 An electrode consisting of Cr (20 nm) and Al (500 nm) is deposited directly on the upper cladding of the waveguide. The separation between the electrode and the waveguide is designed to be 3 $\mu$m, which is estimated to be close enough that the electric field extending from the electrode can effectively influence the LNOI waveguide, but far enough that the optical loss is not increased. Figure \ref{fig4}(a) shows the simulation result of the static electric potential performed using a finite element solver, with the voltage applied across the top and bottom electrodes.  The bottom electrode, serving as a ground plane, is made from Cr (10 nm), Au (100nm) and Cr (10 nm).

To demonstrate the electro-optic tuning of the device we apply a DC voltage from 0\,V down to -55\,V to the top electrode of the ring resonator with a radius of 70 $\mu$m. The resonance shifts with the applied voltage as shown in Fig \ref{fig4}(b) corresponding to a EO tunability of 3 pm/V.

\section{Discussion}

The Fabry-Perot transmission measurements were conducted on straight waveguides with inverse tapers at both ends and indicate low propagation loss for this platform.
The overall insertion loss of the waveguides is dominated by mode-mismatch between waveguide and optical fiber, despite the significant improvement of provided the inverse tapers.
Given that the straight waveguides measured have identical dimensions to the waveguides used in the ring resonators and were fabricated on the same chip, the propagation loss in the rings are concluded to be equally low loss.

The demonstrated ring resonators are designed to be strongly overcoupled, increasing their 3 dB resonance bandwidth (and, conversely, reducing their $Q$-factor).  A 300 nm gap in the bus waveguide to microring coupling region provides strong overcoupling. The potential of the  nanofabrication process used in this work \cite{Krasnokutska:18} could be further extended to photonic components including grating couplers and compact directional couplers.

The $Q$-factor measurements show that it is possible to achieve small and high performance microring resonators for the TM mode---critical for electro-optic and nonlinear applications.  Meanwhile, the TE mode bending losses significantly limit the $Q$-factor of the smaller radius microring resonators; however, increasing the ring radius leads to a substantial increase in $Q$-factor. It demonstrated that the TM mode can achieve a smaller bend losses than the TE mode, as the index contrast of the TE fundamental mode is less than that of the TM fundamental mode, as verified by both our in-house mode solver, and by the $Q$-factor simulations conducted in Lumerical Mode for the 30 \um ring. 

It was found that the theoretically predicted results for the microring resonators demonstrated in this paper are in a good agreement with the experimental results (Fig. \ref{fig3}). The deviation for $n_{g}$ is less than $2\%$ leading to precise agreement between the designed and measured FSRs for different ring geometries. Using 350 nm deep ribs, a small TM bend radius was achieved to enable 30 \um TM microring resonators with an FSR of 5.7 nm.  This result is competitive with other high-index contrast leading platforms, such as SiN and AlN, 

For comparison, we report in Table \ref{tab:EO comparison} a summary of experimental results on resonant wavelength tuning.
It can be seen the tunable ring resonators have been realized in a multitude of photonic platforms. Silicon has reported very high EO tuning with large FSR \cite{Li:11}. More recent work has shown good performance in hybrid Si on LN, although it requires extra fabrication steps \cite{Chen:13, Chen:14}. While using AlN has so far resulted in limited tunability \cite{Jung:14}, LNOI photonics presents a promising approach to tunable ring resonators \cite{Wang:18, Guarino:2007bv, Siew:cx}. The results presented in this work combine good EO tunability with simple fabrication process of $Z$-cut LNOI single mode waveguides, which are readily compatible with other single mode photonic components and and will enable future low-loss and tunable filtering in LNOI.

\begin{center}
\begin{table*}
\caption{\label{tab:EO comparison} Comparison between different types of tunable rings.}
\begin{tabular}{ | l | l | l | l | l |} 
\hline
Material & Radius of a ring (\um) & Q-factor & FSR (nm) & EO tuning (pm/V) \\ 
\hline
SOI PN junction \cite{Li:11} & 7.5 & 8000 & 12.6 & 26\\
\hline
SOI on LN with integrated electrodes \cite{Chen:13} & 15 & 11500 & 7.15 & 12.5 \\
\hline 
SOI on LN \cite{Chen:14} & 15 & 14000 & 7.15 & 3.3\\
\hline
AlN \cite {Jung:14} & 60 & 500000 & n.a. & 0.18 \\ 
\hline
$X$-cut LNOI \cite{Wang:18} & n.a. & 50000 & n.a. & 7\\
\hline
$Z$-cut LNOI \cite{Guarino:2007bv} & 100 & 4000 & 1.66 & 1.05 \\ 
\hline
$Z$-cut LNOI \cite{Siew:cx} & 50 & 2800 & 3.2 & 2.15\\
\hline
$Z$-cut LNOI (this work) & 70 & 7500 & 2.5 & 3 \\
\hline 
\end{tabular}
\end{table*}
\end{center}

\section{Conclusion}
We have analyzed in detail the performance of large FSR microring resonators in $Z$-cut LNOI, fabricating rings of varying radii and reporting their characterization for TE and TM polarizations. The demonstrated advanced fabrication enables minimal separation (300\,nm) between monolithically defined adjacent features, whilst maintaining smooth waveguide sidewalls. We have verified that the optical characteristics of the fabricated microring resonators correspond well with the design and simulation.
We have further demonstrated 3pm/V EO tuning of a 70~$\mu$m radius microring.
These results will precede more complex photonic devices in LNOI, ranging from precise filtering with multistage microring resonators to electro-optically tunable devices.\\

\noindent\textbf{Funding}\\
Australian Research Council Centre for Quantum Computation and Communication Technology CE170100012; Australian Research Council Discovery Early Career Researcher Award, Project No. DE140101700; RMIT University Vice-Chancellors Senior Research Fellowship.

\noindent\textbf{Acknowledgments}\\
We thank Jochen Schr\"{o}der for discussions. This work was performed in part at the Melbourne Centre for Nanofabrication in the Victorian Node of the Australian National Fabrication Facility (ANFF) and the Nanolab at Swinburne University of Technology. The authors acknowledge the facilities, and the scientific and technical assistance, of the Australian Microscopy \& Microanalysis Research Facility at RMIT University.

%

\end{document}